\documentclass[twocolumn]{aastex6}

\usepackage{verbatim}
\usepackage{aas_macros}
\usepackage{hyperref}
\usepackage{natbib}
\usepackage{graphicx}
\usepackage{epsfig}
\usepackage{amsmath}
\usepackage[english]{babel}

\defcitealias{CervantesSodi13}{CS+13}

\shorttitle{Do Low Surface Brightness galaxies host stellar bars?}
\shortauthors{Cervantes Sodi \& S\'anchez Garc\'ia}

\begin{document}

\title{Do Low Surface Brightness galaxies host stellar bars?}

\author{Bernardo Cervantes Sodi \altaffilmark{1} \& Osbaldo S\'anchez Garc\'ia \altaffilmark{1}}
\altaffiltext{1}{Instituto de Radioastronom\'ia y Astrof\'isica, Universidad Nacional Aut\'onoma de M\'exico, Campus Morelia, A.P. 3-72, C.P. 58089 Michoac\'an, M\'exico; b.cervantes@crya.unam.mx}

\begin{abstract}
With the aim of assessing if low surface brightness galaxies host stellar bars,
and study the dependence of the occurrence of bars as a function of surface
brightness, we use the Galaxy Zoo 2 dataset to construct a large volume-limited
sample of galaxies, and segregate the galaxies as low and high surface brightness
in terms of their central surface brightness. We find that the fraction of low surface
brightness galaxies hosting strong bars is systematically lower than the one found for high surface brightness galaxies.
The dependence of the bar fraction on the central surface brightness is mostly driven
by a correlation of the surface brightness with the spin and the gas-richness of the galaxies,
showing only a minor dependence on the surface brightness. We also find that the length
of the bars shows a strong dependence on the surface brightness, and although some
of this dependence is attributed to the gas content, even at fixed gas-to-stellar mass ratio,
high surface brightness galaxies host longer bars than their low surface brightness counterparts,
which we attribute to an anticorrelation of the surface brightness with the spin.

\end{abstract}


\section{Introduction}

Low surface brightness galaxies (LSBs) are faint galaxies that emit less light per
unit area than a typical bright spiral galaxy. 
The term LSB typically refers to disk galaxies with $B$-band central surface brightnesses fainter
than $\mu_0(B)  \geq  $ 22.7 mag arcsec$^{-2}$ \citep{Freeman70}. In this paper we adopt
the commonly used criterion $\mu_0(B) \geq $ 22.0 mag arcsec$^{-2}$ \citep{Impey01,Boissier03,Zhong08}
to define our sample of LSBs. Despite some of them being intrinsically
bright, due to their faint central surface brightness, they were difficult to detect, but
in the last two decades a large amount of research has enlightened our understanding of
this kind of galaxies. Although intrinsically faint, observational studies by \cite{Impey&Bothun97}
and \cite{Oneil&Bothun00} suggest that a significant fraction of the galaxy population
are LSBs, making these galaxies the major recipients of baryonic matter in the Universe.
In general, the span range of physical parameters of LSBs is the same as the conventional
Hubble sequence and are not restricted to be low mass dwarf galaxies \citep{McGaugh95}.
The main difference between LSBs and their high surface brightness (HSB) counterparts is
their low stellar surface mass density that produces the low surface brightness that characterizes
them. The typical rotation curve of LSBs do not rise as steeply as the rotation curves of HSBs of
similar luminosity \citep{deBlok01,Swaters10}, but they are surprisingly flat and extend to large radii
implying that they are strongly dark matter dominated at all radii, with mass models
indicating that the dark matter halos of LSBs are less dense than those of HSBs \citep{deBlok96,deBlok97,
deBlok01}.

In the literature, LSBs are regarded as unevolved galaxies due to their low star formation rates \citep{
vanderHulst93,vanZee97,Wyder09,vanderHoek00,Schombert11}, their high gas fractions for their stellar masses,
and high HI total masses \citep{Burkholder01,Oneil04,Huang14} and low metallicities
\citep{deBlok98,vanderHulst98_b,deNaray04}, which suggest that these kind of galaxies follow
different evolutionary tracks than HSBs.

The existence of LSBs is usually explained as this kind of galaxies being formed in high angular momentum dark
matter halos. If the specific angular momentum of baryons is conserved and set equal to that of
the dark matter halo \citep{Fall80}, then the disk scalelength is regulated by the dark matter halo spin. In this way,
the low surface brightness, which is a direct consequence of the low stellar mass density of the disk, is
specified by the $\lambda$ spin parameter of the dark matter halo, making LSBs systems from the high tail of the
galaxy spin distribution \citep{Dalcanton97,Jimenez98,Mo98,Jimenez03}, with $\lambda > 0.05$ \citep{Boissier03,Kim13}.
In this context, LSBs are galaxies with sparse disks embedded in dark matter halos which are dynamically dominant at
all radii, hence expected to be stable against disk instabilities \citep{Ostriker73, DeBuhr12, Yurin15, Algorry16}.

\cite{Mihos97}, through analytic arguments and
numerical experiments, found that LSBs are stable against both local and global disk instabilities. On a work entirely
focused on the formation and evolution of bars in LSBs, \cite{Mayer04} using high-resolution simulations of LSBs
embedded in cold dark matter halos including models with dominant gaseous components, find that the halo-to-disk
mass ratio within the disk radius is the main factor determining if a galaxy is able to develop a bar. For their LSBs models
with $\lambda \sim 0.065$, a high baryonic fraction of 10 percent is required in order to grow bars, which is more than
double the baryon fraction estimated for LSBs \citep{Hernandez98}. \cite{Mayer04} also report that when LSBs are
able to form bars, their typical sizes are smaller than the halo scale radius and both, gaseous and stellar bars evolve
into forming bulge-like structures in a few gigayears.

If LSBs are galaxies formed in halos with high values of the spin parameter, the result by \cite{Long14} directly
applies to them. Using numerical simulations of isolated galaxies, they demonstrate that the growth
of stellar bars in spinning dark matter halos is strongly suppressed, with the bar growth in strength and size
being increasingly quenched in systems with $\lambda \geq 0.03$, because the angular momentum transfer
from the disk to the halo is reduced. In this same line, \citet{CervantesSodi13}
(henceforth \citetalias{CervantesSodi13}), using a volume limited sample
of galaxies from the Sloan Digital Sky Survey (SDSS), and using an order of magnitude estimate for the spin
parameter \citep{Hernandez06}, found that the fraction of galaxies hosting strong bars was a strong function
of $\lambda$, with strong bars preferentially found in galaxies with low to intermediate values of the spin
parameter, while weak bars presented systematically high values of $\lambda$.

With LSBs expected to be stable against bar formation, it is not surprising to find that bars are rare in LSBs.
In the study by \cite{McGaugh94}, of 36 LSBs only one shows features of a strong bar, and form
\cite{Impey96} catalog, of 516 only 19 are classified as barred galaxies. More recently, \cite{Honey16}
found that of a sample of 856 LSB galaxies extracted from the SDSS, only 8.3 per cent have bars,
and through a near-infrared image study, they conclude that the range of bar parameters such as bar
length and ellipticities are similar to those found in HSBs.

In the present work, we select a large volume-limited sample of galaxies from the SDSS, with visual classification
from the Galaxy Zoo 2 project, to study the fraction of barred galaxies as a function of central surface brightness,
with the aim of establishing if the bar fraction is different between LSBs and HSBs, if the difference is due entirely on
the surface brightness of the galaxies, and if the length of the bars detected on both samples is statistically
different. The sample description is
detailed in Section 2, our general results and discussion are presented in Section 3. Lastly, we summarize our
general conclusions in Section 4.

\begin{figure*}[]
  \begin{center}
    \includegraphics [width=0.7\hsize]{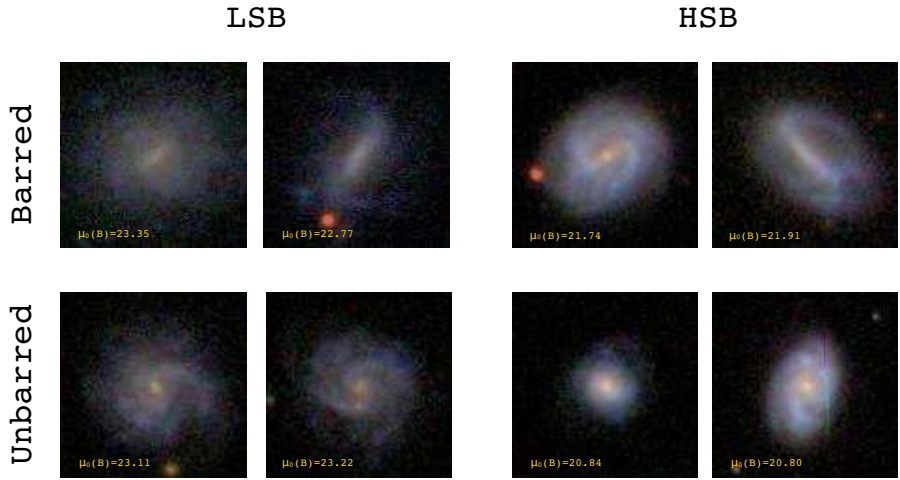}   
  
    \caption{Combined $g+r+i$ color images of examples of LSBs (left panels) and HSBs
    (right panels) classified as barred (upper panels) and unbarred (lower panels). Galaxies $\mu_{0}(B)$ values
    are shown for each case.}
  \label{images}
  \end{center}
\end{figure*}

\section{Data}

The sample of galaxies used in the present study comes from the Galaxy Zoo 2 project.
The Galaxy Zoo 2 \citep[GZ2;][]{Willett13} is an on-line citizen science project where
morphological classification of 304 122 galaxies drawn from the Sloan Digital Sky Survey
Data Release 7 (SDSS DR7) were conducted. Like an extension of its predecessor,
the Galaxy Zoo (\citealt{Lintott08}, \citeyear{Lintott11}), the classification was conducted
following a multi-step decision tree, containing 13 classification tasks and 36 possible
answers guiding the volunteers in order to classify each galaxy by providing a composite
color image (in the \textit{g}, \textit{r} and \textit{i} bands) for the galaxies in the sample.

Given that the present work focuses on barred disk galaxies, we look for galaxies classified
as disk galaxies viewed face-on for a direct identification of the bar.  We impose a strict
classification criterion requesting a minimum of 20 votes for each relevant task; ie, disk,
smooth, face-on. To classify a galaxy as a disk, face-on system we request a vote fraction
$\geq0.6$ (ie, $p_{disk} \ \geq 0.6$,  \ $p_{not  \  edgeon} \ \geq 0.6$),
and to identify barred galaxies we adopt the threshold $p_{bar} \ \geq 0.6$; with all other
galaxies regarded as unbarred ones. A similar threshold of $p_{bar} \ = 0.5$ was adopted
by  \cite{Masters12}, where they showed it to be reliable to identify strong bars; in this sense
we regard weakly barred galaxies as unbarred, and our barred fraction refers to strongly
barred galaxies exclusively.
We restrict our analysis to a volume-limited sample with $0.02<z<0.04865$
and absolute magnitude $M_{r}<-19.7$.

All photometric data was extracted from the SDSS DR7 \citep{Abazajian09}
and corrected using the standard Galactic extinction corrections
\citep{Schlegel98} and a small $k$-correction \citep{Blanton07}. Stellar masses 
were obtained from the MPA/JHU SDSS database\footnote{http://www.mpa-garching.mpg.de/SDSS/}. 
Our final sample consist of 10 430 face-on disk galaxies
segregated into 7 851 unbarred and 2 579 barred galaxies.

To define our subsample of LSBs we use the relation by \citet{Zhong08} to find the central
surface brightness, given by

\begin{equation}
\mu_{0} = m + 2.5 log_{10}(2\pi a^{2})+ 2.5log_{10}(q) - 10log_{10}(1+z)
\end{equation}

where $m$ is the apparent magnitude, $a$ is the disk scale length (in arcsec), $q=b/a$
is the axis ratio and $z$ is the redshift. The third and fourth term of Equation 1 are due to the
correction by inclination and cosmological dimming effects. Once estimated $\mu_{0}$
in the $r$ and $g$ bands, we make use of the transformation equation by \citet{Smith02}
to get $\mu_{0}$ in the $B-$band:

\begin{equation}
\mu_{0}(B) = \mu_{0}(g) + 0.47(\mu_{0}(g) - \mu_{0}(r))+ 0.17.
\end{equation}

Having estimated the central surface brightness on the $B-$band, we are now able to define our
subsample of LSBs as those with $\mu_0(B) \geq $ 22.0, resulting in a total of 4 484 galaxies.

To study the joint dependence of the bar fraction on $\mu_{0}(B)$ and the gas mass fraction $M_{\mathbf{HI}}/M_{*}$,
we use estimates of HI mass from a recently
published catalog by \cite{HIcatalog} who provide $M_{\mathrm{HI}}/M_{*}$ for half a million galaxies in the SDSS,
obtained by applying an artificial neural network based on a sample of $\sim$13,000 SDSS galaxies with H I mass detections
from the Arecibo Legacy Fast Arecibo L-band Feed Array (ALFALFA). This H I estimate shows no systematic trend
with parameters such as stellar mass, color or star formation rate. When using this estimate, our sample is reduced to 8 865 galaxies.

Finally, to explore if the length of the bars found in LSBs is different to that measured on HSBs,
we use the \cite{Hoyle11} catalog, that provides bar lengths for a total of 3 150 galaxies.
As part of the GZ2 project, bar lengths are measured by volunteers though a Google maps
interface, by drawing a line extending the length of the bar in question, requiring at least 3 measurements
per galaxy. By matching our sample with this catalog we obtain a total of 1 686 galaxies with
$L_{\mathrm{bar}}$ measurements, of them 1 362 have H I mass estimates from \cite{HIcatalog}.

Figure \ref{images} shows random example images of LSBs and HSBs with and without bars in our sample.

\section{Results}
\label{Results}

\begin{figure}[]
\label{distributions}
\centering
\begin{tabular}{c}
\includegraphics[width=0.45\textwidth]{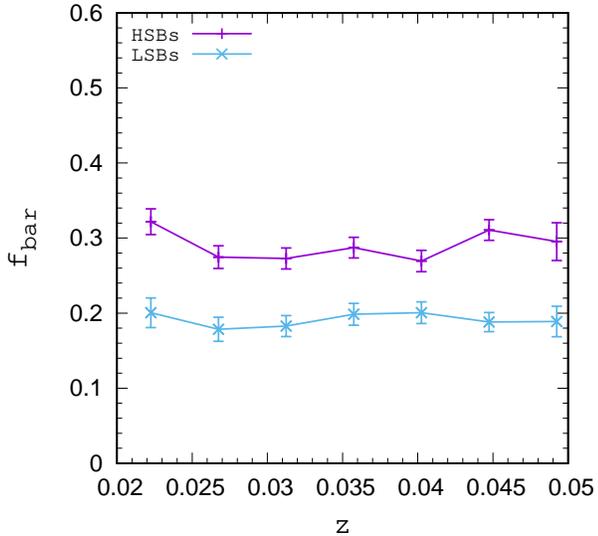} 
\end{tabular}
\caption{Bar fraction as a function of redshift for LSBs and HSBs in our sample.
}\label{fbarVsZ}
\end{figure}

\begin{figure}[]
\label{distributions}
\centering
\begin{tabular}{cc}
\includegraphics[width=0.45\textwidth]{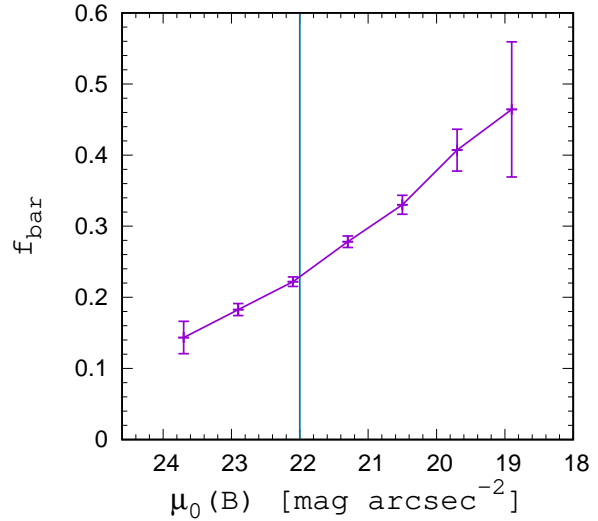}  \\
\includegraphics[width=0.45\textwidth]{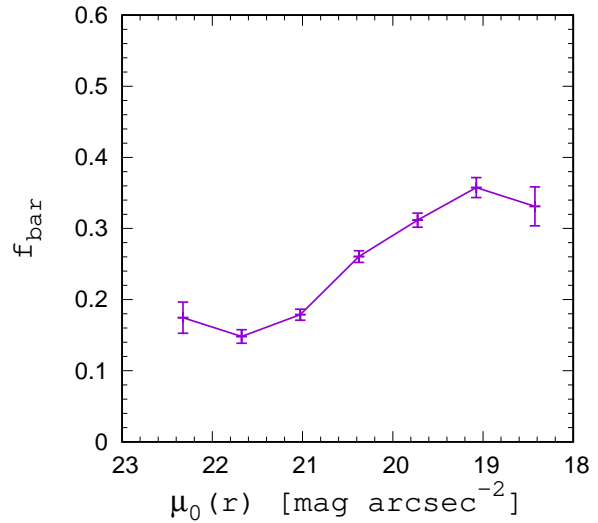} 
\end{tabular}
\caption{Bar fraction as a function of $\mu_{0}(B)$  (\textit{top panel})
and $\mu_{0}(r)$ (\textit{bottom panel}) for strong bars. The vertical line in the top panel denotes the value that segregates
LSBs to the left and HSBs to the right.
}\label{fbarVsmu}
\end{figure}

The fraction of barred galaxies as a function of the limited redshift range of our sample is presented in Figure \ref{fbarVsZ},
where error bars denote the estimated 1$\sigma$ confidence intervals based on the bootstrapping resampling
method. This convention will be kept in all subsequent figures.
The bar fraction for the whole sample is 25$\%$, in good agreement with \cite{Lee12},
who report 24$\%$ of the galaxies in their sample presenting strong bars. We remind the reader
that in the present work, we focus exclusively on strong bars, given that weak bars are not resolved in
this redshift range \citep{Masters11}. This is not an issue regarding strong bars as we did not find any
dependence of the bar fraction on the size of the galaxies in pixels.
For HSBs the bar fraction is close to 30$\%$, also in good agreement with previous works \citep{Eskridge00,
Laurikainen04,Marinova07,Lee12}. As can be appreciated
from the figure, the bar fraction for LSBs is only 20$\%$, systematically lower than the value
estimated for HSBs. The difference of the bar fraction for LSBs and HSBs is due to the dependence of $f_\mathrm{bar}$ on
the central surface brightness, as can be appreciated in Figure \ref{fbarVsmu} top panel, that shows the dependence of $f_\mathrm{bar}$
on $\mu_{0}(B)$, with the vertical line denoting the value that segregates LSBs to the left from HSBs to the right. The corresponding figure
using the natural SDSS photometric $r$-band surface brightness is presented in Figure \ref{fbarVsmu} bottom panel, with the same general
trend, an increase of $f_\mathrm{bar}$ with increasing $\mu_{0}(r)$.

As mentioned in the introduction, this dependence of the bar fraction on the surface brightness was already pointed out by several works
using more limited samples \citep{McGaugh94, Honey16}, and predicted by \citetalias{CervantesSodi13} in the case LSB represent
the tail of high spinning galaxies in the general galaxy population. To test this hypothesis, we used the model proposed by 
\citetalias{CervantesSodi13} to estimate the spin parameter $\lambda$ as defined by \cite{Peebles71} 
$\lambda = L  E^{1/2}/G M^{5/2}$ , where $E$ is the total energy, $M$ the mass,
and $L$ the angular momentum of the configuration. The model by \citetalias{CervantesSodi13} includes a dark matter halo with a
truncated isothermal density profile responsible for establishing a rigorously flat rotation curve along the whole disk, 
and a disk with an
exponential surface density profile of the form $\Sigma(r)=\Sigma_{0} e^{-r/R_{d}}
$, with  $\Sigma_{0}$ the central surface density and $R_{d}$ the disk scalelength. 
Circular velocities ($V_{d}$) are assigned through a Tully-Fisher relation \citep{Pizagno07} and a
disk-to-halo mass ratio ($f_{d}$) is adopted following the prescription by \cite{Gnedin07} to finally express $\lambda$ as:

\begin{equation}
\label{GeneralLambda}
\lambda =  \left({\sqrt{2}}\over{G} \right) {f_{d}} R_{d}  V_{d}^{2}M_{d}^{-1}
\end{equation}

\begin{figure}[]
\label{distributions}
\centering
\begin{tabular}{cc}
\includegraphics[width=0.5\textwidth]{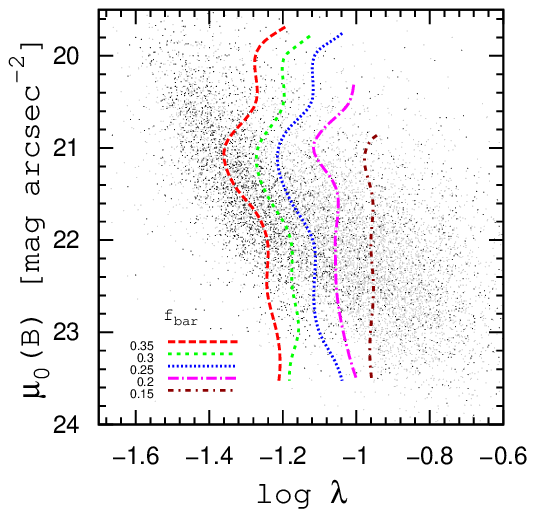} \\
\includegraphics[width=0.45\textwidth]{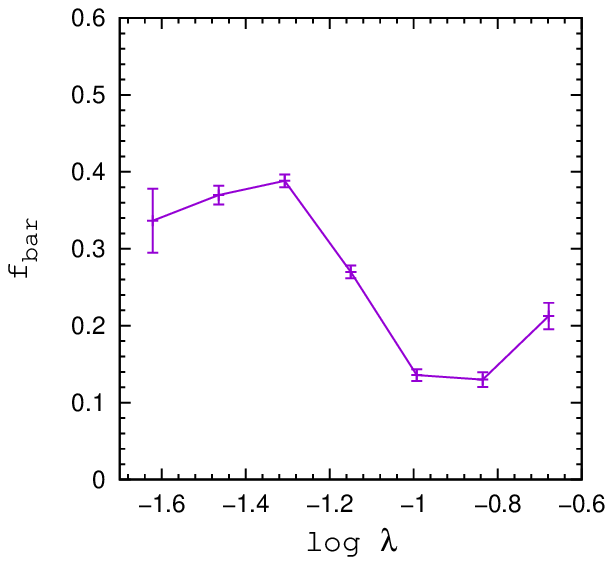} 
\end{tabular}
\caption{\textit{Top panel:} Bar fraction $f_{\mathrm{bar}}$ isocontours in the $\mu_{0}(B)$ vs. $\lambda$ plane.
Contours denote regions of constant $f_{\mathrm{bar}}$ in the range $0.15 < f_{\mathrm{bar}} < 0.35$.
Gray dots represent unbarred galaxies, black dots represent barred ones.
\textit{Bottom panel:} Bar fraction as a function of the $\lambda$ spin parameter.
}\label{fbVsSpin}
\end{figure}

For a detailed description of the model, we refer
the reader to \citetalias{CervantesSodi13}. Figure \ref{fbVsSpin} top panel clearly shows an anticorrelation between $\mu_{0}(B)$ and $\lambda$,
confirming our hypothesis that LSBs form in high spinning systems. In this sense, the increase of $f_\mathrm{bar}$ with increasing
$\mu_{0}(B)$ is a natural outcome, given the previous result by \citetalias{CervantesSodi13}. 
In the same figure we include isocontours that denote regions of constant $f_{\mathrm{bar}}$ obtained by dividing the
parameter space into 10 x 10 bins, applying a spline kernel to get a smooth transition, and requiring at least
5 galaxies per bin to estimate the bar fraction. A clear trend of increasing $f_{\mathrm{bar}}$ with decreasing $\lambda$
is noticeable with no dependence on $\mu_{0}(B)$ at fixed $\lambda$. From this figure is apparent that the dependence of
$f_{\mathrm{bar}}$ on $\mu_{0}(B)$ comes from the marked correlation between $\mu_{0}(B)$ and $\lambda$.
For completeness we show in Figure \ref{fbVsSpin} bottom panel
the bar fraction as a function of $\lambda$, recovering the same behavior reported by \citetalias{CervantesSodi13} (see their figure 1b), with a decrease
of the bar fraction with increasing $\lambda$. This result
is encouraging given that in the sample used in the present work the bar identification is made by amateur citizen
scientists, while in the sample employed by \citetalias{CervantesSodi13} the classification is preformed by professional astronomers \citep{Lee12}.

\begin{figure*}
\label{distributions}
\centering
\begin{tabular}{cc}
\includegraphics[width=0.4\textwidth]{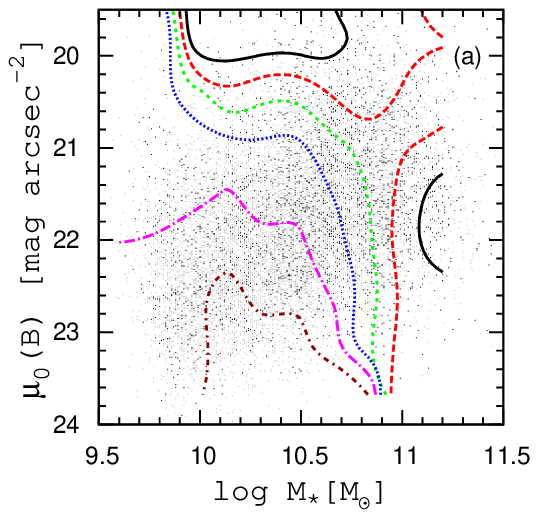} & \includegraphics[width=0.4\textwidth]{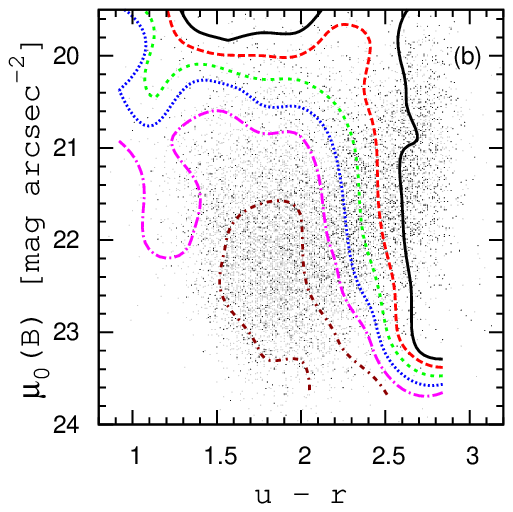} \\
\includegraphics[width=0.4\textwidth]{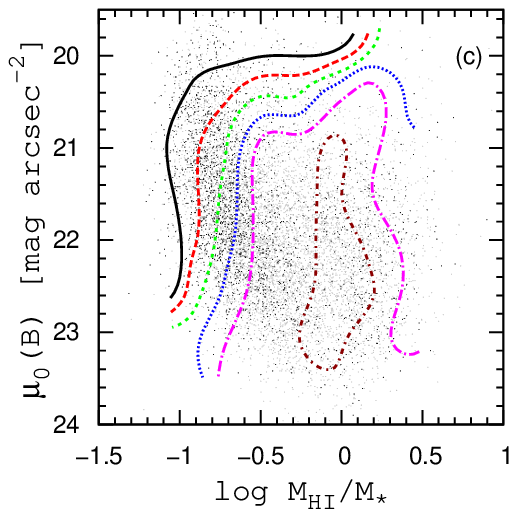} & \includegraphics[width=0.4\textwidth]{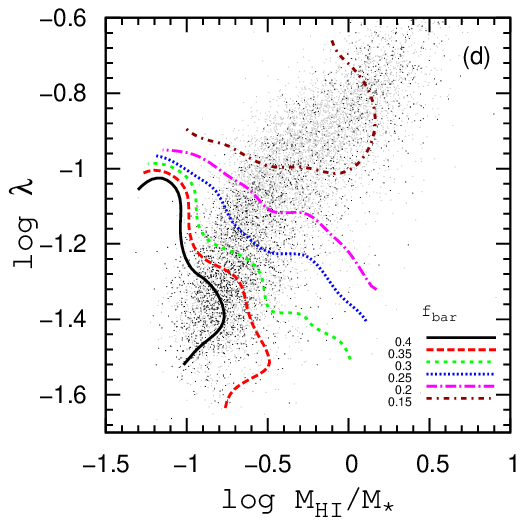}
\end{tabular}
\caption{Bar fraction $f_{\mathrm{bar}}$ isocontours in the $\mu_{0}(B)$ vs. $M_{\mathrm{*}}$ (a), $\mu_{0}(B)$ vs. $u-r$ (b),
$\mu_{0}(B)$ vs. $M_{\mathrm{H I}}/M_{\mathrm{*}}$ (c), and $\mu_{0}(B)$ vs. $\lambda$ (d) planes.
Contours denote regions of constant $f_{\mathrm{bar}}$ in the range $0.15 < f_{\mathrm{bar}} < 0.40$. Gray dots represent unbarred
galaxies, black dots represent barred ones.
}\label{2D}
\end{figure*}

As several studies have previously pointed out \citep{Masters12, Oh12, Skibba12, CervantesSodi13, Gavazzi15}, the bar fraction
is a strong function of stellar mass, with increasing $f_\mathrm{bar}$ as $M_\mathrm{*}$ increases. Likewise, LSBs tend
to be less massive than their HSBs counterparts \citep{Galaz11}, which could explain why $f_\mathrm{bar}$ is lower for LSBs.
In Figure \ref{2D}a, we explore the co-dependence of the bar fraction on $\mu_\mathrm{0}(B)$ and $M_\mathrm{*}$. The first
thing to notice on the figure is a general increase of $\mu_\mathrm{0}(B)$ with increasing $M_\mathrm{*}$. Secondly,
for galaxies with $M_\mathrm{*}>10^{10.75}$, the iso-contours denoting constant $f_\mathrm{bar}$ show a strong
dependence on stellar mass, while almost no dependence on $\mu_\mathrm{0}(B)$. For less massive galaxies,
a co-dependence with $\mu_\mathrm{0}(B)$ appears, with $f_\mathrm{bar}$ decreasing with increasing
central surface brightness. This kind of bimodality has already being detected, for instance \cite{Nair10}
showed that the bar fraction presents a minimum at $M_\mathrm{*}\sim10^{10.2}$, a bimodality which is
also seen in the stellar population.

The fraction of barred galaxies is also sensitive to color, $f_\mathrm{bar}$ being highest for red galaxies \citep{Nair10,
Masters11, Lee12, Oh12}. Figure \ref{2D}b shows that $f_\mathrm{bar}$ depends exclusively on $u-r$ and is independent of
$\mu_\mathrm{0}(B)$ for red galaxies with $u-r > 2$, but for blue galaxies the dependence of $f_\mathrm{bar}$ on $\mu_\mathrm{0}(B)$
is noticeable, stressing the presence of the aforementioned bimodality.

\begin{figure}[]
\label{distributions}
\centering
\begin{tabular}{cc}
\includegraphics[width=0.45\textwidth]{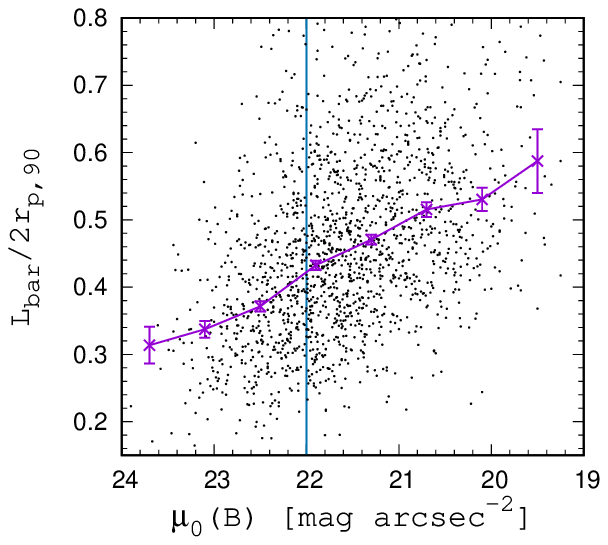}  \\
\includegraphics[width=0.45\textwidth]{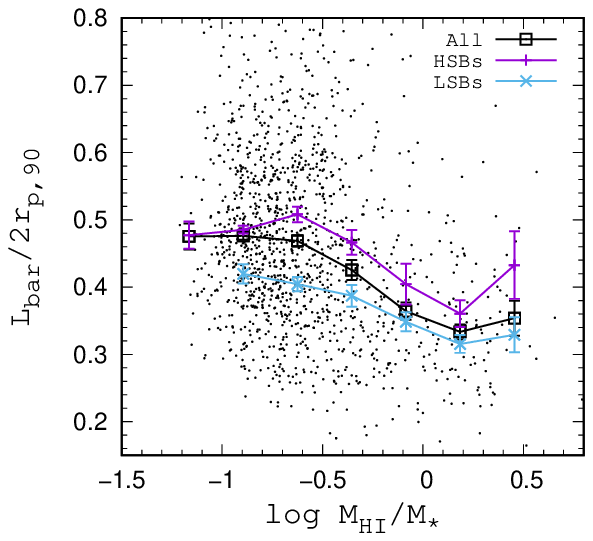}  \\
\includegraphics[width=0.45\textwidth]{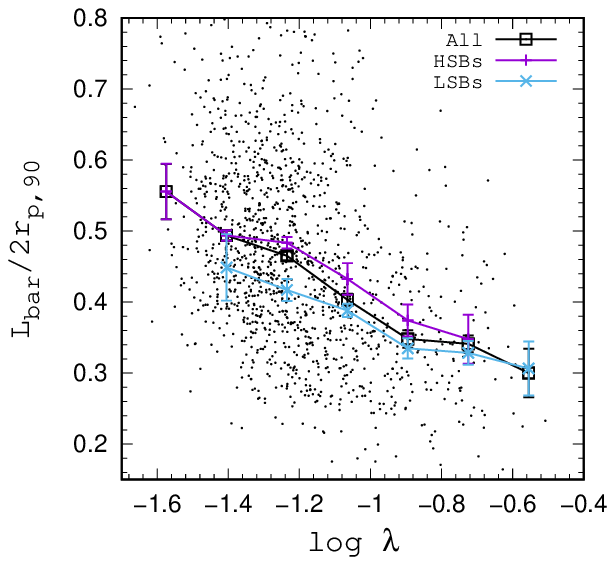}  
\end{tabular}
\caption{\textit{Top panel:} normalized bar length as a function of $\mu_{0}(B)$,
the vertical line denotes the value that segregates
LSBs to the left and HSBs to the right. \textit{Middle panel:} normalized bar length as a
function of $M_\mathrm{H I}/M_\mathrm{*}$ for the full sample (black solid line), as
well as the sample segregated in LSBs (dash-dotted blue line) and HSBs
(dashed violet line). \textit{Bottom panel:} normalized bar length as a function
of $\lambda$, line coding the same as in middle panel.
}\label{Lbar}
\end{figure}

Using samples of galaxies from the SDSS with H I gas mass estimates from
ALFALFA \citep{Giovanelli05}, \cite{Masters12} and \cite{CervantesSodi17}
showed the inhibiting effect gas has in the formation of bars.
Using the H I mass estimate from \cite{HIcatalog}, we explore in 
Figure \ref{2D}c the co-dependence of $f_\mathrm{bar}$ on $\mu_\mathrm{0}(B)$ and $M_\mathrm{H I}/M_\mathrm{*}$,
finding that, for galaxies fainter than $\mu_0(B) = $ 21.0 mag arcsec$^{-2}$,
at fixed $M_\mathrm{H I}/M_\mathrm{*}$ there is almost no dependence on $\mu_\mathrm{0}(B)$. This figure
indicates that the dependence of $f_\mathrm{bar}$ on the surface brightness of the galaxies is mostly driven
by the correlation present between surface brightness and gas richness. In this sense, given that LSBs are
intrinsically gas-rich galaxies \citep{Minchin04, Pustilnik11, Du15}, the low fraction of barred LSBs is at least partially driven
by the inhibiting effect the gas has in the formation of bars \citep{Masters12, CervantesSodi17},
which is more evident in LSBs than in HSBs given that they present a wider range of
$M_\mathrm{H I}/M_\mathrm{*}$ values. Figure \ref{2D}c also shows that for galaxies brighter than
$\mu_0(B) = $ 21.0 mag arcsec$^{-2}$, surface brightness may play a role in the likelihood of galaxies
hosting bars, even for gas-rich systems.

Finally, Figure \ref{2D}d shows a clear correlation between the $\lambda$ spin parameter and
the gas mass fraction, as previously reported by \cite{CervantesSodi09, Huang12}. From the contours
almost entirely perpendicular to the distribution of points in the plane, we can conclude that the dependence
of $f_\mathrm{bar}$ with $\lambda$ is as strong as the dependence of $f_\mathrm{bar}$ with
the gas fraction. Considering that at fixed $\lambda$ (Figure \ref{fbVsSpin}a)
and $M_\mathrm{H I}/M_\mathrm{*}$ (Figure \ref{2D}c),
the bar fraction is independent of $\mu_\mathrm{0}(B)$, we conclude that the dependence of $f_\mathrm{bar}$
on $\mu_\mathrm{0}(B)$ is driven by these other two co-dependences.

Given that we also count with bar length measurements from \cite{Hoyle11}, we explored the dependence of the
bar length ($L_\mathrm{bar}$), normalized to the optical size of the galaxy defined as two times the
$r-$band Petrosian radius 90 (2$r_\mathrm{p,90}$), on $\mu_\mathrm{0}(B)$ (\ref{Lbar} top panel),
finding a clear trend of increasing bar size with increasing $\mu_\mathrm{0}(B)$. 

In Figure \ref{2D}c we learned that the low fraction of barred LSBs is mostly driven by the large gas mass fraction of
this type of galaxies. Taking into account the results obtained through numerical experiments by \cite{Villa-Vargas10}
and \cite{Athana13}, where bars are prevented to form in gas-rich galaxies, and in the cases where
they are able to form, they form later and are weaker than in gas-poor galaxies, the increase of bar length with
increasing surface brightness observed in Figure \ref{Lbar} top panel, could be due to the trend between
$\mu_\mathrm{0}(B)$ and $M_\mathrm{H I}/M_\mathrm{*}$, and not directly attributed to the the surface
brightness. To disentangle this dependence of the bar length with $\mu_\mathrm{0}(B)$ and $M_\mathrm{H I}/M_\mathrm{*}$
we turn to Figure \ref{Lbar} middle panel, where we show the normalized bar length as a function of
$M_\mathrm{H I}/M_\mathrm{*}$ for the full sample (black solid line) as well as for the case of the
sample segregated in HSBs and LSBs (purple and blue lines respectively). As expected, as we
move from gas-poor to gas-rich systems, the bar length decreases for all cases, the full sample as well
as for the HSBs and LSBs considered separately. Interestingly, at fixed gas-mass fraction, the HSBs have
systematically longer bars than the LSBs, indicating that the bar length is not independent of
surface brightness. We also explored the dependence of the bar length on $\lambda$ in Figure \ref{Lbar} lower panel,
finding a steadily decrease of the bar length with increasing spin, with HSBs having again, systematically longer
bars than their LSBs counterparts.

\section{Conclusions}

We compiled a large volume-limited sample of galaxies drawn from the SDSS-DR7, and studied
the dependence of the bar fraction on the surface brightness of galaxies. We find that the
fraction of galaxies hosting strong bars among LSBs ($\sim$ 20\%) is systematically lower than the fraction
found for HSBs ($\sim$ 30\%), a natural consequence derived from the dependence of $f_\mathrm{bar}$
with $\mu_\mathrm{0}(B)$ found in the present work, with the fraction
of barred galaxies increasing with increasing central surface brightness.

By exploring the co-dependence of the bar fraction with $\mu_\mathrm{0}(B)$ and other physical
parameters of the galaxies in the sample, such as stellar mass, color, and gas-to-stellar mass
fraction, we conclude that the increase of $f_\mathrm{bar}$  with increasing $\mu_\mathrm{0}(B)$
is mostly driven by an anti-correlation of $\mu_\mathrm{0}(B)$ with $\lambda$ and $M_\mathrm{H I}/M_\mathrm{*}$.
In this sense, the low bar fraction of LSBs is primarily due to their H I gas-richness when compared with
their HSBs counterparts, and the inhibiting effect the gas has in the formation of bars, as previously
pointed out by numerous works \citep{Masters12, CervantesSodi17,Kim17}, plus the fact
that LSBs are high spinning galaxies, making their disks more sparse and less self-gravitating, and hence
less prone to global instabilities, a result in accordance with the previous result by  \citetalias{CervantesSodi13}.

While the presence of bars in galaxies seems to be fairly independent of the surface brightness of
galaxies at a fixed $M_\mathrm{H I}/M_\mathrm{*}$, the bar length shows a stronger correlation
with $\mu_\mathrm{0}(B)$ than with $M_\mathrm{H I}/M_\mathrm{*}$, and at fixed gas-to-stellar mass
ratio, HSBs have systematically larger bars than LSBs, a trend in good agreement with the theoretical results
by \cite{Mayer04}, who found that the bars formed in simulated LSBs are shorter than the ones
formed in HSBs and that these bars quickly evolve to pseudobulges.
This correlation between the normalized bar length and $\mu_\mathrm{0}(B)$ is mostly driven
by the anticorrelation between $\mu_\mathrm{0}(B)$ and $\lambda$, with decreasing bar length with
increasing $\lambda$, showing only a minor dependence on $\mu_\mathrm{0}(B)$ at fixed $\lambda$.
Our result  concerning  the bar length goes in line with
the findings by \citetalias{CervantesSodi13}, who reports that weak bars are preferentially hosted by
high spinning galaxies (LSBs), while strong bars are more commonly found in galaxies with low-to-intermediate values
of $\lambda$ (HSBs).

\acknowledgments

	The authors acknowledge financial support through PAPIIT project IA103517 from DGAPA-UNAM.
	The work of O. S\'anchez Garc\'ia is supported by a CONACYT scholarship.
	The authors also acknowledge the thorough reading of the original manuscript by the
	anonymous referee, as helpful in reaching a clearer and more complete final version.
    Funding for the SDSS and SDSS-II has been provided by the Alfred P. Sloan Foundation,
    the Participating Institutions, the National Science Foundation, the U.S. Department of
    Energy, the National Aeronautics and Space Administration, the Japanese
    Monbukagakusho, the Max Planck Society, and the Higher Education Funding Council
    for England. The SDSS Web Site is http://www.sdss.org/. The SDSS is managed by the
    Astrophysical Research Consortium for the Participating Institutions. The Participating
    Institutions are the American Museum of Natural History, Astrophysical Institute Potsdam,
    University of Basel, University of Cambridge, Case Western Reserve University,
    University of Chicago, Drexel University, Fermilab, the Institute for Advanced Study, the
    Japan Participation Group, Johns Hopkins University, the Joint Institute for Nuclear Astrophysics,
    the Kavli Institute for Particle Astrophysics and Cosmology, the Korean Scientist Group,
    the Chinese Academy of Sciences (LAMOST), Los Alamos National Laboratory,
    the Max-Planck-Institute for Astronomy (MPIA), the Max-Planck-Institute for Astrophysics (MPA),
    New Mexico State University, Ohio State University, University of Pittsburgh,
    University of Portsmouth, Princeton University, the United States Naval Observatory,
    and the University of Washington.

\end{document}